\documentclass[12pt]{article}
\usepackage{amssymb, amsmath, amsfonts,latexsym, verbatim, amsthm,
mathrsfs}

\title{Topological Factors Derived From Bohmian Mechanics}
\author{
Detlef D\"urr\footnote{Mathematisches Institut der Universit\"{a}t
       M\"{u}nchen, Theresienstra{\ss}e 39, 80333 M\"{u}nchen, Germany.
       E-mail: duerr@mathematik.uni-muenchen.de},
Sheldon Goldstein\footnote{Departments of Mathematics, Physics and
       Philosophy, Hill Center, Rutgers, The State University of New
       Jersey, 110 Frelinghuysen Road, Piscataway, NJ 08854-8019, USA.
       E-mail: oldstein@math.rutgers.edu},
James Taylor\footnote{Center for Talented Youth, Johns Hopkins
University,
McAuley Hall, Suite 400, 5801 Smith Ave, Baltimore, MD 21209, USA. E-
mail:
       jostylr@member.ams.org},\\
Roderich Tumulka\footnote{Mathematisches Institut,
       Eberhard-Karls-Universit\"at, Auf der Morgenstelle 10, 72076
       T\"ubingen, Germany. E-mail:
       tumulka@everest.mathematik.uni-tuebingen.de},
\ and Nino Zangh\`\i\footnote{Dipartimento di Fisica dell'Universit\`a
       di Genova and INFN sezione di Genova, Via Dodecaneso 33, 16146
       Genova, Italy. E-mail: zanghi@ge.infn.it}
}
\date{January 9, 2006} % {\today}

\newcommand{\z}[1]{{#1}}
\newcommand{\zz}[1]{{#1}}
\newcommand{\x}[1]{{#1}}

\newtheorem{assertion}{Assertion}

\newtheorem{defn}{Definition}
{\bf}{\it}

\newcommand{\dud}[2]{\ensuremath{\frac{\partial {#1}}{\partial {#2}}}}

\newcommand{\rvarn}[1]{\ensuremath{\mathbb{R}^{#1}}}

\newcommand{\sdud}[2]{\ensuremath{\frac{d{#1}}{d{#2}}}}

\newcommand{\gp}{\ensuremath{\psi}}
\newcommand{\gD}{\ensuremath{\Delta}}
\newcommand{\gr}{\ensuremath{\pi}}
\newcommand{\re}{\ensuremath{\mathbb{R}}}
\newcommand{\se}{Schr\"odinger's equation}
\newcommand{\sch}{Schr\"odinger}
\newcommand{\qpotc}[1]{\ensuremath{\frac{\hbar^2}{2m_{#1}}}}

\newcommand{\qmu}[1]{\ensuremath{\frac{\hbar}{m_{#1}}}}

\newcommand{\defi}{\ensuremath{:=}}

\newcommand{\seq}[1]{\ensuremath{i \hbar \dud{\gp}{t} = - \sum_{k =
1}^{#1} \qpotc{k} \gD_{k} \gp + V \gp   }}

\newcommand{\im}{\ensuremath{\mathrm{Im}}}
\newcommand{\inpr}[2]{\ensuremath{({#1}, {#2})}}
\newcommand{\dd}{\ensuremath{d}}

\newcommand{\nrd}{\ensuremath{{}^N\mspace{-1.0mu}\rvarn{\dd}}}

\newcommand{\wf}{wave function}

\newcommand{\covspa}{\ensuremath{\widehat{\gencon}}}
\newcommand{\proj}{\ensuremath{\gr   }}
\newcommand{\gencon}{\ensuremath{\mathcal{Q}}}
\newcommand{\fund}[1]{\ensuremath{\pi_1 ({#1})}}
\newcommand{\deckg}{\ensuremath{Cov(\covspa, \gencon)}}
\newcommand{\deckt}{\ensuremath{\sigma}}

\newcommand{\concov}{\ensuremath{\gamma_{\deckt}}}

\newcommand{\conc}[1]{\ensuremath{\gamma_{#1}}}
\newcommand{\cmplx}{\ensuremath{\mathbb{C}}}
\newcommand{\deck}[1]{\ensuremath{\deckt_{#1}}}

\newcommand{\baseq}{\ensuremath{q}}
\newcommand{\covq}{\ensuremath{\hat{q}}}

\newcommand{\canq}{\ensuremath{\boldsymbol{q}}}

\newcommand{\spins}{\ensuremath{\cmplx^{2s +1}}}
\newcommand{\sspa}{\ensuremath{W}}

\newcommand{\hb}{\ensuremath{\hbar}}
\newcommand{\bm}{Bohmian mechanics}

\newcommand{\Sect}[1]{Section {#1}}

\newcommand{\wh}[1]{\ensuremath{\widehat{#1}}}

\newcommand{\fiber}{covering fiber}

\newcommand{\limts}{}
\newcommand{\concav}{\ensuremath{\gamma}}
\newcommand{\sidecom}[1]{}
\newcommand{\herm}{Hermitian}

\newcommand{\id}{\ensuremath{\mathrm{Id}}}

\newcommand{\RRR}{\mathbb{R}} % real numbers
\newcommand{\CCC}{\mathbb{C}} % complex numbers
\newcommand{\ZZZ}{\mathbb{Z}} % integers
\newcommand{\Q}{\gencon} % configuration space
\newcommand{\electric}{\boldsymbol{E}} % electric field strength
\newcommand{\magnetic}{\boldsymbol{B}} % magnetic field strength
\newcommand{\cylinder}{\mathcal{C}} % cylinder in the AB effect
\newcommand{\rawH}{-\tfrac{\hbar^2}{2} \gD + V} %
\newcommand{\holonomy}{\ensuremath{h}} % holonomy endomorphism
\newcommand{\closedcurve}{\alpha}
\newcommand{\opencurve}{\beta}

\newcommand{\nrtre}{{}^N \RRR^3}
\newcommand{\covr}{\hat{r}}
\begin{document}
\maketitle

\begin{abstract}
We derive for Bohmian mechanics topological factors for quantum systems
with a multiply-connected configuration space $\Q$. These include
nonabelian factors corresponding to what we call holonomy-twisted
representations of the fundamental group of $\Q$. We employ wave
functions
on the universal covering space of $\Q$. As a byproduct of our
analysis, we
obtain an explanation, within the framework of Bohmian mechanics, of the
fact that the wave function of a system of identical particles is either
symmetric or anti-symmetric.

\medskip

Key words: topological phases, multiply-connected configuration
spaces, Bohmian mechanics, universal covering space

\noindent MSC (2000):
\underline{81S99}, %(general quantum mechanics and quantization)
81P99, %(foundations of quantum mechanics),
81Q70. %(holonomy in quantum theory)
PACS: 03.65.Vf, %(phases: geometric; dynamic or topological),
03.65.Ta %(foundations of quantum mechanics)
\end{abstract}

\begin{center}\textit{Dedicated to Rafael Sorkin on the occasion of
his 60th birthday}\end{center}
%\tableofcontents

\section{Introduction} \label{sec:intro}

We \zz{consider here} a novel approach towards topological effects in quantum
mechanics. These effects arise when the configuration space $\Q$ \zz{of a
quantum system} is a
multiply-connected Riemannian manifold and involve \emph{topological
factors} forming a representation
(or holonomy-twisted representation) of the fundamental group $\fund
{\Q}$ of $\Q$. Our approach is based on Bohmian mechanics \cite
{Bohm52, Bell66, DGZ92, survey, DGZ96, Gol01}, a version of quantum
mechanics with particle trajectories. The use of Bohmian paths allows
a derivation of the link between homotopy and quantum mechanics that
is essentially different from derivations based on \zz{path integrals.}

The topological factors we derive are equally relevant and applicable
in orthodox quantum mechanics, or any other version of quantum
mechanics. Bohmian mechanics, however, provides a sharp mathematical
justification of the dynamics with these topological factors that is
absent in the orthodox framework. Different topological factors give
rise to different Bohmian dynamics, and thus to different quantum
theories, for the same configuration space $\Q$ (whose metric we
regard as incorporating the ``masses of the particles''), the same
potential, and the same
value space of the wave function.

The motion of the configuration in a Bohmian \zz{system of $N$ distinguishable
particles} can be regarded as \z{corresponding to} a dynamical system in
the configuration space $\Q = \RRR^{3N}$, defined by a time-dependent
vector field $v^{\psi_t}$ on $\Q$ which in turn is defined, by the Bohmian
law of motion, in terms of $\psi_t$. We are concerned here with the
analogues of the Bohmian law of motion \zz{when} $\Q$ is, instead
of $\RRR^{3N}$, an arbitrary Riemannian manifold.\footnote{Manifolds will
throughout be assumed to be Hausdorff, paracompact, connected, and
$C^\infty$.  They need not be orientable.}  The main result is that, if
$\Q$ is multiply connected, there are several such analogues: several
dynamics, which we will describe in detail, corresponding to different
choices of the topological factors.

\zz{It is easy to overlook} the multitude of dynamics by focusing too much
 on just one, the simplest one, which we will define in
 \Sect{\ref{sec:immediate}}: the \emph{immediate generalization} of the
 Bohmian dynamics from $\RRR^{3N}$ to a Riemannian manifold, or, as we
 shall briefly call it, the \emph{immediate Bohmian dynamics}. \zz{Of the
 other kinds of Bohmian dynamics, the simplest involve phase factors
 associated with non-contractible loops in $\Q$,} forming a
 character\footnote{By a \emph{character} of a group we refer to what is
 sometimes called a unitary multiplicative character, i.e., a one-
 dimensional unitary representation of the group.} of the fundamental group
 $\fund{\Q}$. In other cases, \zz{the topological factors} are given by
 matrices \zz{or endomorphisms}, forming a \z{unitary representation of
 $\fund{\Q}$} or, in the case of a vector bundle, a holonomy-twisted
 representation (see the end of \Sect{\ref{s:bundlevalued}} for the
 definition).  As we shall explain, the dynamics of bosons is an
 ``immediate'' one, but not the dynamics of fermions (except when using a
 certain \zz{not entirely natural} vector bundle). The Aharonov--Bohm
 effect can be regarded as an example of a non-immediate dynamics on the
 \zz{accessible region of 3-space.}

It is not obvious what ``other kinds of Bohmian dynamics''
should mean. We will investigate one approach here, while \zz{others}
will be studied
in forthcoming works. The present approach is
based on
considering wave functions $\psi$ that are defined not on the
configuration space $\Q$ but on its universal covering space
$\covspa$. We then \zz{investigate} which kinds of periodicity conditions,
relating the values on different levels of the covering fiber by a
topological factor, will ensure that the Bohmian velocity vector field
associated with $\psi$ is projectable from $\covspa$ to $\Q$.  This is
carried out in \Sect{\ref{sec:covering}} for scalar wave functions and
in \Sect{\ref{s:periodic2}} for wave functions with values in a
complex vector space (such as \z{a} spin-space) or a complex vector
bundle.
In the case of vector bundles, we derive a novel kind of topological
\z{factor}, given by  a holonomy-twisted \z{representation}  of $\fund
{\Q}$.

The notion that multiply-connected spaces give rise to different
topological factors
is not new.  The most common approach is based on path integrals and
began
   largely with the work of Schulman \cite{S68,Sch71} and Laidlaw
   and DeWitt \cite{DL71}; see \cite{Sch81} for details.
Nelson \cite{Nel85} derives the topological phase factors for
scalar wave functions from stochastic mechanics. There is also the
current
algebra approach of Goldin, Menikoff, and Sharp \cite{GMS81}.

\section{Bohmian Mechanics in Riemannian manifolds}
\label{sec:bm}
\label{sec:immediate}

\renewcommand{\dd}{\ensuremath{3}}

Bohmian mechanics can be formulated by appealing only to the
Riemannian structure $g$ of the configuration $\Q$ space of a
physical system:   the state of the system in \bm\ is given by the
pair $(Q,
\gp)$; $Q \in \Q $ is
the configuration of \zz{the} system and $\psi$ is a
(standard quantum mechanical) \wf\ on the configuration space $\Q$,
taking values in some \emph{Hermitian vector space} $\sspa$, i.e., a
finite-dimensional complex vector space endowed with a
positive-definite Hermitian (i.e., conjugate-symmetric and
sesqui-linear) inner product $(\,\cdot\,,\,\cdot\,)$.

The state of
the system changes according to the guiding equation and \se{} \cite
{DGZ92}:

\begin{subequations} \label{ie:re}
\begin{align}
    \sdud{Q_t}{t} &= v^{\psi_t}(Q_t)\label{ie:rbe} \\
    i \hb \dud{\psi_t}{t}  &= - \tfrac{\hb^2}{2} \gD \psi_t
    + V \psi_t
    \, ,\label{ie:rse}
\end{align}
\end{subequations}
where the Bohmian velocity vector field $v^{\psi}$ associated \zz{with} the
\wf\ $\psi$ is
\begin{equation} \label{ie:rbv}
    v^{\psi} \defi \hb\, \im \frac{\inpr{\psi}{\nabla
    \psi}}{\inpr{\psi}{\psi}}.
\end{equation}
In the above equations $\Delta$ and $\nabla$ are,
respectively, the Laplace-Beltrami operator and the gradient on the
configuration space equipped with this Riemannian structure;
$V$ is the potential function with values \z{given by}  Hermitian
matrices (endomorphisms of $\sspa$).
Thus, given $\gencon$, $\sspa$, and $V$, we \z{have specified} a Bohmian
dynamics, the \emph{immediate Bohmian dynamics}.\footnote{Since the
law of motion  involves a derivative
    of $\psi$, the merely measurable functions in $L^2(\Q)$
    will of course not be \z{adequate} for defining trajectories.
    However, we will leave aside the \z{question,}  from which dense
    subspace of $L^2(\Q)$ \z{should one}  choose $\psi$. For a
discussion of the global existence question of Bohmian trajectories
in $\RRR^{3N}$, see \cite{BDGPZ95,TT04}.}

The empirical agreement between \bm\ and standard quantum mechanics is
grounded in equivariance \cite{DGZ92,DGZ03}.  In \bm, if the
configuration
is initially random and distributed according to $|\gp_0|^2$, then the
evolution is such that the configuration at time $t$ will be
distributed according to $|\gp_t|^2$.  This property is called \z{the}
equivariance of the $|\gp|^2$ distribution.  It follows from comparing
the transport equation arising from \eqref{ie:rbe}
\begin{equation}\label{continuity}
    \dud{\rho_t}{t} =- \nabla \cdot (\rho_t v^{\gp_t})
\end{equation}
for the distribution $\rho_t$ of the configuration $Q_t$, where
$v^\psi = (v_1^\psi, \ldots, v_N^\psi)$, to \z{the quantum continuity
equation}
\begin{equation}\label{dpsi2dt}
    \dud{|\gp_t|^2}{t} =- \nabla \cdot (|\gp_t|^2 v^{\gp_t}),
\end{equation}
which is a consequence of \sch's equation \eqref{ie:rse}. A
rigorous proof of equivariance requires showing that almost all (with
respect to the $|\gp|^2$ distribution) solutions of \eqref{ie:rbe}
exist for all times. This was done in \cite{BDGPZ95,TT04}.  A more
comprehensive introduction to \bm\ may be found in \cite{Gol01,
survey, DGZ96}.

An important example \zz{(with, say, $\sspa=\CCC$) is that}  of several particles moving in \z{a}
Riemannian manifold $M$, a \z{possibly} curved physical space. Then the
configuration space for $N$ \zz{distinguishable} particles is $\gencon \defi
M^{N}$.  Let the masses of the particles be $m_i$ and the metric of
$M$ be $g$. Then the relevant metric on $M^N$ is
\[%\begin{equation*}
    g^N(v_1 \oplus \cdots \oplus v_N, w_1 \oplus \cdots \oplus w_N )
    \defi \sum\limts_{i=1}^N m_i g(v_i,w_i).
\]%\end{equation*}
Using $g^N$ allows us to write \eqref{ie:rbv} and \eqref{ie:rbe}
instead of the equivalent equations
\begin{equation}\label{e:bohm02}
    \sdud{\boldsymbol{Q}_{k}}{t} = \qmu{k} \im
    \frac {\inpr{\psi}{\nabla_{k} \psi}}{ \inpr{\psi}{\psi}}
    (\boldsymbol{Q}_1,\ldots, \boldsymbol{Q}_N),
    \quad  k= 1,\ldots,N
\end{equation}
\begin{equation}\label{e:sch02}
    \seq{N},
\end{equation}
where $\boldsymbol{Q}_k$, the $k^{th}$ component of $Q$, lies in $M$,
and
$\nabla_k$ and $\gD_k$ are the gradient and the Laplacian with
respect to
$g$, acting on the $k^{th}$ factor of $M^N$.
\zz{Another important} example \cite{DL71} is that of $N$ identical
particles in $
\rvarn{3}$,   for which the natural configuration space
is the \z{set $\nrd$ of} all $N$-element subsets of $\rvarn{3}$,
\begin{equation}
     \nrd := \{S| S \subseteq \rvarn{3}, |S| = N\} \,,
\end{equation}
which inherits a Riemannian metric from $\rvarn{3}$. \zz{Spin is
incorporated by choosing  for $\sspa$ \z{a}  suitable spin space
\cite{Bell66}.  For one particle moving in \rvarn{3}, we may take \sspa\ to
be} a complex,
irreducible representation space of $SU(2)$, the universal covering
group\footnote{The universal covering space of a Lie group is again a
    Lie group, the \emph{universal covering group}. It should be
    distinguished from another group also called the \emph{covering
      group}: the group $Cov(\covspa,\gencon)$ of the covering (or deck)
    transformations of the universal covering space $\covspa$ of a
    manifold $\Q$, which will play an important role later.} of the
rotation group $SO(3)$.  If it is the spin-$s$ representation then
$\sspa = \spins$.

More generally, we can consider a Bohmian dynamics for \wf s
taking values in a complex vector bundle $E$ over the Riemannian
manifold \gencon. That is, the value space then depends on the
configuration, and \wf s become sections of the vector
bundle. Such a case occurs for identical particles with spin $s$,
where the
bundle $E$ of spin spaces over the configuration space $\Q = \nrd$
consists of the $(2s+1)^N$-dimensional
spaces
\begin{equation}\label{spinbundle}
    E_q = \bigotimes_{\canq \in q} \cmplx^{2s+1} \,, \quad q \in \Q\,.
\end{equation}
For a detailed discussion of this bundle, of why this is the right
bundle, and of the notion of a tensor product over an arbitrary index
set, \x{see  \cite{topid2}}.

We introduce now some notation and terminology.

\begin{defn}
    A \emph{\herm\ vector bundle}, \z{or} \emph{\herm\ bundle}, \zz{over
    $\Q$} is a finite-dimensional complex vector bundle $E$ \zz{over $\Q$}
    with a connection and a positive-definite, Hermitian local inner
    \zz{product $(\,\cdot\,,\,\cdot\,)=(\,\cdot\,,\,\cdot\,)_q$ on $E_q$,
    the fiber of $E$ over $q\in\Q$,} which is parallel.
\end{defn}

Our bundle, the one of which $\psi$ is a section, will always be a \herm\
bundle. Note that since a \herm\ bundle consists of a vector bundle and a
connection, it can be nontrivial even if the vector bundle is trivial:
namely, if the connection is nontrivial. The \emph{trivial \herm\ bundle}
$\Q \times \sspa$, in contrast, consists of the trivial vector bundle with
the trivial connection, whose parallel transport $P_\opencurve$\zz{, in
general a unitary endomorphism from $E_q$ to $E_{q'}$ for $\opencurve$ a path
from $q$ to $q'$,} is always the identity on $\sspa$.
%Note that trivial connections are flat.
The \z{case of a $W$-valued function}  $\psi : \Q \to \sspa$
corresponds to the trivial \herm\
bundle $\Q \times \sspa$.

The global inner product on the Hilbert space of \wf s is the
local inner product integrated against the Riemannian volume measure
associated with \x{the metric $g$ of $\Q$,}
\[
    \langle \phi, \psi \rangle = \int_{\gencon} dq \, (\phi(q),
    \psi(q))\,.
\]
The Hilbert space equipped with this inner product, denoted
$L^2(\gencon,E)$, contains the square-integrable, measurable (not
necessarily smooth) sections of $E$ modulo equality almost everywhere.
The covariant derivative $D\psi$ of a section $\psi$ is an ``$E$-valued
1-form,'' i.e., a section of $\CCC T\gencon^* \otimes E$ (with $T\Q^*
$ the
cotangent bundle), while we write $\nabla \psi$ for the section of $\CCC
T\gencon \otimes E$ metrically equivalent to $D\psi$.  The potential
$V$ is now a self-adjoint section of the endomorphism
bundle $E \otimes E^*$ acting on the vector bundle's fibers.
The equations defining the Bohmian dynamics are, {\em mutatis
mutandis}, the same equations \eqref{ie:re} \zz{and \eqref{ie:rbv}} as before.

We wish to introduce now further Bohmian dynamics beyond the immediate one.
To this end, we will consider wave functions on \covspa{}, the universal
covering space of $\Q$. This idea is rather standard in the literature on
quantum mechanics in multiply-connected spaces \cite{DL71, D72, LM77,
Mor92, HM96}.  However, the complete \zz{specification} of the
possibilities that we give in \Sect{\ref{s:periodic2}} includes some,
corresponding to what we call \emph{holonomy-twisted representations} of
$\fund{\Q}$, that have not \z{yet} been considered.  \zz{Each possibility}
has locally the same Hamiltonian $\rawH$, with the same potential $V$, and
each possibility is equally well defined and equally reasonable.  While in
orthodox quantum mechanics it may seem \zz{more or less axiomatic that} the
configuration space $\Q$ is the space on which $\psi_t$ is defined, $\Q$
appears in Bohmian mechanics also in another role: as the space in which
$Q_t$ moves. It is therefore less surprising from the Bohmian viewpoint,
and easier to accept, that $\psi_t$ is defined not on $\Q$ but on
$\covspa$.  In the next section all \wf s will be complex-valued; in
\Sect{\ref{s:periodic2}} we shall consider wave functions with
higher-dimensional value spaces.

\section{Scalar  Wave Functions on the Covering Space}
\label{sec:covering}

The motion of the configuration $Q_t$ in \gencon\ is determined by a
velocity vector field $v_t$ on \gencon, which may arise from a wave
function $\psi$ not on \gencon\ but instead on \covspa\ , the
universal covering space of $\Q$, in the
following way:
Suppose we are given a complex-valued map $\gamma$ on the covering
group
$\deckg$, \zz{$\gamma: \deckg \to \cmplx$, and suppose}
that a wave function $\psi: \covspa \to \cmplx$ satisfies the
\emph{periodicity condition associated with the topological factors
    $\gamma$}, i.e.,
\begin{equation} \label{e:percon}
    \psi (\deckt \covq) = \concov \psi(\covq)
\end{equation}
for every $\covq \in \covspa$ and $\deckt \in \deckg$.
For \eqref{e:percon} to be
possible for a $\psi$ that does not identically vanish, $\gamma$ must
be a representation of the covering group, as was first emphasized in
\cite{D72}. To see this, let $\deck1$, $\deck2 \in \deckg$. Then we
have the following equalities
\begin{equation}\label{e:cccargument}
\conc{\deck1 \deck2} \psi (\covq)
= \psi(\deck1 \deck2 \covq)
= \conc{\deck1} \psi(\deck2 \covq)
= \conc{\deck1}\conc{\deck2} \psi(\covq).
\end{equation}
We \z{thus obtain}  the fundamental relation
\begin{equation} \label{e:ccc}
    \conc{\deck1 \deck2} = \conc{\deck1}\conc{\deck2},
\end{equation}
establishing \z{(since $\gamma_{\id} =1$)}  that $\gamma$
is a representation.

Let \fund{\gencon, \baseq} denote the \emph{fundamental group \zz{of $\Q$}
at a point} \baseq{} and let $\proj$ be \zz{the} covering map (a local
diffeomorphism) $\proj: \covspa \to \gencon$, also called the projection
(the \emph{\fiber} for $\baseq\ \in \gencon$ is the set
$\proj^{-1}(\baseq)$ of points in \covspa\ that project to \baseq\ under
$\proj$).  The 1-dimensional representations of the covering group are,
\z{via the canonical isomorphisms $\varphi_{\covq}: \deckg \to
\fund{\gencon, q},\ \covq\in
\proj^{-1}(\baseq)$,}
%\x{\footnote{$\varphi_{\covq}(\sigma)$ is the homotopy
%equivalence class corresponding to a path from $\covq$ to
%$\sigma^{-1}\covq$.}}
in canonical correspondence with the 1-dimensional
representations of any fundamental group \fund{\gencon, q}: \z{The
different} isomorphisms \z{$\varphi_{\covq},\ \covq\in
\proj^{-1}(\baseq)$,} will \z{transform a representation} of \fund{\gencon,
q} into \z{representations of $\deckg$ that are conjugate. But the}
1-dimensional representations are homomorphisms to the \emph{abelian}
multiplicative group of $\cmplx$ and \z{are} thus invariant under
conjugation.

{}From \eqref{e:percon} it follows that $\nabla \psi(\deckt \covq) =
\gamma_\deckt \, \deckt^* \nabla \psi(\covq)$, where $\deckt^*$ is the
(push-forward) action of $\deckt$ on tangent vectors, using that
$\deckt$ is an isometry. Thus, the velocity field $\hat{v}^{\psi}$ on
\covspa\ associated with $\psi$ according to
\begin{equation}\label{vhatdef}
    \hat{v}^\psi (\covq) := \hbar \, \im \, \frac{ \nabla
    \psi}{ \psi} (\covq)
\end{equation}
is projectable, i.e.,
\begin{equation}\label{vhatprojectable}
    \hat{v}^\psi (\deckt\covq) = \deckt^* \hat{v}^\psi (\covq),
\end{equation}
and therefore gives rise to a velocity field $v^\psi$ on
\gencon,
\begin{equation}
    v^\psi(q) = \proj^* \, \hat{v}^\psi (\covq)
\end{equation}
where $\covq$ is an arbitrary element of $\proj^{-1}(q)$.

If we let $\psi$ evolve according to the Schr\"odinger equation on
\covspa,
\begin{equation}\label{e:sch04}
    i\hbar \frac{\partial \psi}{\partial t}(\covq) = - \tfrac{\hbar^2}
{2} \gD
    \psi(\covq) + \widehat{V}(\covq) \psi(\covq)
\end{equation}
with $\widehat{V}$ the lift of the potential $V$ on $\gencon$, then
the periodicity condition \eqref{e:percon} is preserved by the
evolution, since, according to
\begin{equation}
    i\hbar\frac{\partial \psi}{\partial t}(\deckt \covq)
    \stackrel{\eqref{e:sch04}}{=} -\tfrac{\hbar^2}{2} \gD \psi(\deckt
    \covq) + \widehat{V}(\deckt \covq) \psi(\deckt \covq) =
    -\tfrac{\hbar^2}{2} \gD \psi(\deckt \covq) + \widehat{V}(\covq)
    \psi(\deckt \covq)
\end{equation}
(note the different arguments in the potential), the functions $\psi
\circ \deckt$ and $\gamma_\deckt \psi$ satisfy the same evolution
equation \eqref{e:sch04} with, by \eqref{e:percon}, the same initial
condition, and thus coincide at all times.

\z{Therefore}  we can let the Bohmian configuration $Q_t$ move according
to $v^{\psi_t}$,
\begin{equation}\label{e:bohm04}
    \frac{dQ_t}{dt} = v^{\psi_t}(Q_t) = \hbar\, \proj^* \Bigl( \im\,
    \frac{ \nabla \psi}{
    \psi}\Bigr)(Q_t) = \hbar\, \proj^* \Bigl( \im\,
    \frac{ \nabla \psi}{
    \psi}\Big|_{\covq \in \proj^{-1}(Q_t)} \Bigr).
\end{equation}
One can also view the motion in this way: Given $Q_0$, choose
$\widehat{Q}_0 \in \proj^{-1}(Q_0)$, let $\widehat{Q}_t$ move in
\covspa\ according to $\hat{v}^{\psi_t}$, and set $Q_t =
\proj(\widehat{Q}_t)$. Then the motion of $Q_t$ is independent of the
choice of $\widehat{Q}_0$ in the fiber over $Q_0$, and obeys
\eqref{e:bohm04}.

If, as we shall assume from now on, $|\gamma_\deckt|=1$ for all
$\deckt \in \deckg$, i.e., if $\gamma$ is a \emph{unitary}
representation (in $\cmplx$) or a \emph{character}, then the motion
\eqref{e:bohm04} also has an equivariant probability distribution,
namely
\begin{equation}\label{e:equi04}
    \rho(q) = |\psi(\covq)|^2.
\end{equation}
To see this, note that we have
\begin{equation}\label{projectablepsi2}
    |\psi(\deckt \covq)|^2 \stackrel{\eqref{e:percon}}{=}
    |\gamma_\deckt|^2 |\psi(\covq)|^2 = |\psi(\covq)|^2,
\end{equation}
so that the function $|\psi(\covq)|^2$ is projectable to a function on
\gencon\ which we call $|\psi|^2(q)$ in this paragraph. From
\eqref{e:sch04} we have \z{that}
\[
    \frac{\partial |\psi_t(\covq)|^2}{\partial t} = - \nabla \cdot
\Bigl(
    |\psi_t(\covq)|^2 \, \hat{v}^{\psi_t}(\covq) \Bigr)
\]
and, by projection, \z{that}
\[
    \frac{\partial |\psi_t|^2(q)}{\partial t} = - \nabla \cdot \Bigl(
    |\psi_t|^2 (q)\, v^{\psi_t}(q) \Bigr),
\]
which coincides with the transport equation for a probability density
$\rho$ on \gencon,
\[
    \frac{\partial \rho_t(q)}{\partial t} = - \nabla \cdot \Bigl(
    \rho_t(q) \, v^{\psi_t}(q) \Bigr).
\]
Hence,
\begin{equation}
    \rho_t(q) = |\psi_t|^2(q)
\end{equation}
for all times if \z{it is} so initially; this is equivariance.

\zz{The relevant} wave functions are those with
\begin{equation}
    \int_\gencon dq \, |\psi(\covq)|^2 = 1
\end{equation}
where the choice of $\covq \in \proj^{-1}(q)$ is arbitrary by
\eqref{projectablepsi2}. The relevant Hilbert space, which we denote
$L^2(\covspa,\gamma)$, thus \z{consists of}  the measurable functions
$\psi$
on $\covspa$ (modulo changes on null sets) satisfying \eqref{e:percon}
with
\begin{equation}
    \int_\gencon dq \, |\psi(\covq)|^2 < \infty.
\end{equation}
It is a Hilbert space with the scalar product
\begin{equation}
    \langle \phi,\psi \rangle = \int_\gencon dq \, \overline{\phi
(\covq)}
    \, \psi(\covq).
\end{equation}
Note that the value of the integrand at $q$ is independent of the
choice of
$\covq \in \proj^{-1}(q)$ since, by \eqref{e:percon} and \z{the fact
that}
$|\gamma_\deckt|=1$,
\[
    \overline{\phi(\deckt \covq)} \, \psi(\deckt \covq) =
    \overline{\gamma_\deckt \, \phi(\covq)} \, \gamma_\deckt \,
    \psi(\covq) = \overline{\phi(\covq)} \, \psi(\covq).
\]

We summarize the results of our reasoning.

\begin{assertion}\label{a:scalar}
    Given a Riemannian manifold $\gencon$ and a smooth function
    $V:\gencon \to \re$, there is a Bohmian dynamics in \gencon\ with
    potential $V$ for each character \concav\ of the fundamental group
    \fund{\gencon}; it is defined by \eqref{e:percon}, \eqref{e:sch04},
    and \eqref{e:bohm04}, where the wave function $\psi_t$ lies in
    $L^2(\covspa,\gamma)$ and has norm one.
\end{assertion}
Assertion~\ref{a:scalar} provides  as many \zz{dynamics} as there are
characters of $\fund{\Q}$ because different characters $\gamma' \neq
\gamma$ always define different dynamics.
\label{is:rem}
In particular,  for the trivial character $\gamma_\deckt =1$, we
obtain the
    immediate dynamics, as defined by \eqref{ie:rbv} and \eqref{ie:re}.

An  important application of Assertion~\ref{a:scalar}  is
    provided by identical particles without spin.  The natural
    configuration space $\nrd$ \z{for}  identical particles has
fundamental group $S_N$, the group of
    permutations of $N$ objects, which possesses two characters, the
    trivial character, $\gamma_\sigma =1$, and the alternating
    character, $\gamma_\sigma = \mathrm{sgn}(\sigma)= 1$ or $-1$
    depending on whether $\sigma \in S_N$ is an even or an odd
    permutation. The Bohmian
    dynamics associated with the trivial character is that of bosons,
    while the one associated with the alternating character is that of
    fermions. However, in a two-dimensional world there would be more
possibilities
    since $\fund{^N \rvarn2}$ is the braid group, whose \zz{generators
    $\sigma_i, \ i=1,\dots,N-1$,} are
a certain subset of  braids that exchange two particles and
satisfy the \zz{defining  relations}
\begin{align*}
\sigma_i \sigma_j &=\sigma_j\sigma_i \quad \hbox{for} \quad i\leq N-3, j \geq i
+ 2 , \\
\sigma_i\sigma_{i+1}\sigma_i &= \sigma_{i+1}\sigma_i \sigma_{i+1} \quad
\hbox{for} \quad i\leq N-2.
\end{align*}
\zz{Thus, a character} of the braid group assigns the same complex number
$e^{i \beta}$ to each generator, and therefore, according to Assertion~\ref
{a:scalar}, \zz{each choice of $\beta$ corresponds to a Bohmian dynamics;}
two-dimensional bosons correspond to $\beta = 0$ and two- dimensional
fermions to $\beta = \pi$.  The particles corresponding to the other
possibilities are usually called \emph{anyons}. They were first suggested
in \cite{LM77}, and their investigation began in earnest with \cite{GMS81,
Wi82}. See \cite{Mor92} for some more details and references.

\section{Vector-Valued   Wave Functions on the Covering Space}
\label{s:periodic2}
\label{s:vectorvalued}

\label{s:bundlevalued}

\zz{The analysis of \Sect{\ref{sec:covering}} can be carried over with
little change to the case of vector-valued wave functions, $\psi(q)\in
W$. In this case, however, the topological factors may be given by any
endomorphisms $\Gamma_{\sigma}$ of $W$ that form a representation of
$\deckg$ and need not be restricted to characters, a possibility first
mentioned in \cite{Sch81}, Notes to Section 23.3. Rather than directly
considering this case, we focus instead on one that is a bit more general
and that will require a new sort of topological factor, that of wave
functions that are sections of a vector bundle.}  The topological factors
\zz{for this case} will be expressed as \emph{periodicity sections}, i.e.,
parallel unitary sections of the endomorphism bundle indexed by the
covering group \z{and} satisfying a certain composition law, or,
equivalently, as \emph{holonomy-twisted representations} of $\fund{\Q}$.

If $E$ is a vector bundle over \gencon, then the lift of $E$, denoted
by $\wh{E}$, is a vector bundle over \covspa; the fiber space at
\covq\ is defined to be the fiber space of $E$ at $\baseq$,
$\wh{E}_{\covq} \defi E_{\baseq}$, where $\baseq = \proj(\covq)$.  It
is important to realize that with this construction, it makes sense to
ask whether $v \in \wh{E}_{\covq}$ is equal to $w \in \wh{E}_{\covr}$
whenever $\covq$ and $\covr$ are elements of the same covering fiber.
Equivalently, $\wh{E}$ is the pull-back of $E$ through $\proj: \covspa
\to \Q$.  As a particular example, the lift of the tangent bundle of
\gencon\ to \covspa\ is canonically isomorphic to the tangent bundle
of \covspa.  Sections of $E$ or $E\otimes E^*$ can be lifted to
sections of $\wh{E}$ respectively $\wh{E} \otimes \wh{E}^*$.

If $E$ is a \herm\ vector bundle, then so is $\wh{E}$. The wave
function $\psi$ that we consider here is a section of $\wh{E}$, so
that $\psi($\covq$)$ is a vector in the $\covq$-dependent Hermitian
vector space $\wh{E}_{\covq}$. $V$ is a section of the bundle $E
\otimes E^*$, i.e.,
$V(q)$ is an element of $E_q \otimes E_q^*$. To indicate that every
$V(q)$ is a Hermitian endomorphism of $E_q$, we say that $V$ is a
\z{Hermitian section} of $E \otimes E^*$.

Since $\psi(\deckt \covq)$ and $\psi(\covq)$ lie in the same space
$E_q = \wh{E}_{\covq}= \wh{E}_{\deckt \covq}$, a periodicity condition
can be of the form
\begin{equation}\label{e:percon06}
    \psi(\deckt \covq) = \Gamma_\deckt(\covq) \, \psi (\covq)
\end{equation}
for $\deckt \in \deckg$, where $\Gamma_\deckt(\covq)$ is an
endomorphism $E_q \to E_q$.  By the same argument as in
\eqref{e:cccargument}, the condition for \eqref{e:percon06} to be
possible, if $\psi(\covq)$ can be any element of $\wh{E}_{\covq}$, is
the composition law
\begin{equation}\label{e:compo06}
    \Gamma_{\deck1 \deck2}(\covq) = \Gamma_{\deck1} (\deck2 \covq) \,
    \Gamma_{\deck2} (\covq).
\end{equation}
Note that this law differs from the one $\Gamma(\covq)$ would satisfy
if it were a representation, which reads $\Gamma_{\sigma_1 \sigma_2}
(\covq) = \Gamma_{\sigma_1} (\covq) \, \Gamma_{\sigma_2} (\covq)$, \z
{since
in general $\Gamma (\sigma\covq)$ need not be the same as $\Gamma
(\covq)$} .

For periodicity \eqref{e:percon06} to be preserved under the
Schr\"odinger evolution,
\begin{equation}\label{e:sch06}
    i\hbar \frac{\partial \psi}{\partial t} (\covq) = -\tfrac {\hbar^2}
    {2} \gD \psi(\covq) + \wh{V}(\covq) \, \psi(\covq),
\end{equation}
we need that multiplication by $\Gamma_\deckt (\covq)$ \z{commute}  with
the Hamiltonian. Observe that
\begin{equation}\label{HGamma}
    [H,\Gamma_\deckt]\psi(\covq) = -\tfrac{\hbar^2}{2} (\gD
    \Gamma_\deckt(\covq)) \psi(\covq) - \hbar^2 (\nabla
    \Gamma_\deckt(\covq)) \cdot (\nabla \psi(\covq)) +
    [\wh{V}(\covq),\Gamma_\deckt(\covq)] \, \psi(\covq).
\end{equation}
Since we can choose $\psi$ such that, for any one particular $\covq$,
$\psi(\covq)=0$ and $\nabla \psi(\covq)$ is any element of $\CCC T_
{\covq}
\covspa \otimes E_q$ we like, we must have that
\begin{equation}\label{e:parallel06}
    \nabla \Gamma_\deckt(\covq) =0
\end{equation}
for all $\deckt\in \deckg$ and all $\covq \in \covspa$, \z{i.e., that
$\Gamma_\sigma$ is parallel.}  Inserting this
in \eqref{HGamma}, the first two terms on the right hand side
vanish. Since we can choose for $\psi(\covq)$ any element of $E_q$ we
like, we must have that
\begin{equation}\label{e:commute06}
    [\wh{V}(\covq),\Gamma_\deckt(\covq)]=0
\end{equation}
for all $\deckt\in \deckg$ and all $\covq \in \covspa$.  Conversely,
assuming \eqref{e:parallel06} and \eqref{e:commute06}, we obtain that
$\Gamma_\deckt$ commutes with $H$ for every $\deckt\in \deckg$, so
that the periodicity \eqref{e:percon06} is preserved.

\begin{comment}
In this case, we have for
every $q$ a unitary representation $\Gamma(q)$ of $\deckg$ on $E_q$
that commutes with $V(q)$; in addition, by \eqref{e:parallel06},
$\Gamma$ is parallel (covariantly constant). Another way of viewing
$\Gamma$ is this: Denoting the unitary group of $E_q$ by $U(E_q)$, we
can form the group bundle $U(E)$, whose fiber at $q$ is the group
$U(E_q)$. The sections of this group bundle form an
(infinite-dimensional) group under pointwise multiplication, and the
parallel sections form a (finite-dimensional) subgroup. We say that a
section $A$ of $E \otimes E^*$ \emph{commutes pointwise with $V$}, or
simply \emph{commutes with $V$}, if at every $q \in \gencon$, $A(q)$
and $V(q)$ commute. The parallel sections of $U(E)$ that commute with
$V$ form a subgroup of the group of all parallel sections of $U(E)$,
since commuting with $V$ is inherited by products. $\Gamma$ is, by
\eqref{e:parallel06} and \eqref{e:commute06}, an element of that
subgroup.
\end{comment}

{}From \eqref{e:percon06} and \eqref{e:parallel06} it follows that
$\nabla \psi(\deckt \covq) = (\deckt^* \otimes \Gamma_\deckt (\covq))
\nabla \psi(\covq)$. If every $\Gamma_\deckt(\covq)$ is
\emph{unitary}, as we assume from now on, the velocity field
$\hat{v}^\psi$ on \covspa\ associated with $\psi$ according to
\begin{equation}
    \hat{v}^\psi (\covq) := \hbar \, \im \,
    \frac{(\psi,\nabla\psi)}{(\psi,\psi)} (\covq)
\end{equation}
is projectable, $\hat{v}^\psi(\deckt \covq) = \deckt^*
\hat{v}^\psi(\covq)$, and gives rise to a velocity field $v^\psi$ on
\gencon. We let the configuration move according to $v^{\psi_t}$,
\begin{equation}\label{e:bohm06}
    \frac{dQ_t}{dt} = v^{\psi_t}(Q_t) = \hbar \, \proj^* \Bigl( \im \,
    \frac{(\psi,\nabla \psi)}{(\psi,\psi)} \Bigr) (Q_t).
\end{equation}

\begin{defn}
    Let $E$ be a \herm\ bundle over the manifold $\Q$.  A
    \emph{periodicity section} $\Gamma$ over $E$ is a family indexed by
    $\deckg$ of unitary parallel sections $\Gamma_\sigma$ of $\wh{E}
    \otimes \wh{E}^*$ satisfying the composition law \eqref{e:compo06}.
\end{defn}

Since $\Gamma_\deckt(\covq)$ is unitary, one sees as before that the
probability distribution
\begin{equation}\label{e:equi06}
    \rho(q) = (\psi(\covq),\psi(\covq))
\end{equation}
does not depend on the choice of $\covq \in \proj^{-1}(q)$ and is
equivariant.

As usual, we define for any periodicity section $\Gamma$ the Hilbert
space $L^2(\covspa, \wh{E}, \Gamma)$ to be the set of measurable
sections $\psi$ of $\wh{E}$ (modulo changes on null sets) satisfying
\eqref{e:percon06} with
\begin{equation}
    \int_\gencon dq \, (\psi(\covq),\psi(\covq)) < \infty,
\end{equation}
endowed with the scalar product
\begin{equation}
    \langle \phi, \psi \rangle = \int_\gencon dq \,
    (\phi(\covq),\psi(\covq)).
\end{equation}
As before, the value of the integrand at $q$ is independent of the
choice of $\covq \in \proj^{-1}(q)$.

We summarize the results of our reasoning.

\begin{assertion}\label{a:bundle}
    Given a \herm\ bundle $E$ over the Riemannian manifold $\gencon$ and
    a \z{Hermitian section} $V$ of $E \otimes E^*$, there is a
    Bohmian dynamics for each periodicity section $\Gamma$ commuting
    (pointwise) with $\wh{V}$ \z{(cf. (\ref{e:commute06}))}; it is
defined by \eqref{e:percon06},
    \eqref{e:sch06}, and \eqref{e:bohm06}, where the wave function
    $\psi_t$ lies in $L^2(\covspa, \wh{E}, \Gamma)$ and has norm 1.
\end{assertion}

Every character $\gamma$ of \z{$\deckg$ (or of $\fund{\Q}$)}  defines
a periodicity section
by setting
\begin{equation}\label{Gammagamma2}
\Gamma_\sigma (\covq) := \gamma_\sigma
\id_{\wh{E}_{\covq}}.
\end{equation}
It commutes with every potential $V$.
Conversely, a periodicity section $\Gamma$ that commutes with every
potential must be such that every $\Gamma_\sigma (\covq)$ is a
multiple of the identity, $\Gamma_\sigma (\covq) = \gamma_\sigma
(\covq) \, \id_{\wh{E}_{\covq}}$. By unitarity, $|\gamma_\sigma| =1$;
by parallelity \eqref{e:parallel06}, $\gamma_\sigma (\covq) =
\gamma_\sigma$ must be constant; by the composition law
\eqref{e:compo06}, $\gamma$ must be a homomorphism, and thus a
character.

\z{We briefly indicate} how a periodicity section $\Gamma$
corresponds to
something like a representation of $\fund{\Q}$.  Fix a $\covq \in
\covspa$. \z{Then $\deckg$ can be identified with
$\fund{\Q}=\fund{\Q,\proj(\covq)}$ via $\varphi_{\covq}$.}  Since the
sections $\Gamma_\sigma$ of $\wh{E} \otimes \wh{E}^*$
are parallel, $\Gamma_\sigma(\covr)$ is determined for every $\covr$ by
$\Gamma_\sigma(\covq)$. \z{(Note in particular that the parallel
transport
$\Gamma_\sigma(\tau\covq)$ of $\Gamma_\sigma(\covq)$ from $\covq$ to
$\tau\covq, \tau\in \deckg$, may differ from $\Gamma_\sigma(\covq)
$.)} Thus,
the periodicity section $\Gamma$ is completely determined by the
endomorphisms $\Gamma_\sigma := \Gamma_\sigma(\covq)$ of $E_q$, $
\sigma \in
\deckg$, which satisfy the composition law
\begin{equation}\label{twistedrep}
    \Gamma_{\sigma_1 \sigma_2} = \holonomy_{\closedcurve_2} \Gamma_
{\sigma_1}
    \holonomy_{\closedcurve_2}^{-1} \Gamma_{\sigma_2}\,,
\end{equation}
\z{where $\closedcurve_2$}  is any loop in $\Q$ based at $\proj(\covq)
$ whose
lift starting at $\covq$ leads to $\sigma_2 \covq$, and
$\holonomy_{\closedcurve_2}$ is the associated holonomy endomorphism of
$E_q$.  Since \eqref{twistedrep} is not the composition law
$\Gamma_{\sigma_1 \sigma_2} = \Gamma_{\sigma_1} \Gamma_{\sigma_2}$ of
a representation, the $\Gamma_\sigma$ \z{form, not}  a representation of
$\fund{\Q}$, \z{ but}  what we call a \emph{holonomy-twisted
representation}.

\zz{The situation  where the wave function
assumes values in a fixed Hermitian space $\sspa$,} instead of a bundle,
corresponds to the trivial \herm\ bundle $E = \gencon \times \sspa$
(i.e., with the trivial connection, for which parallel transport is
the identity on $\sspa$). Then, parallelity \eqref{e:parallel06}
implies that $\Gamma_\deckt (\covr) = \Gamma_\deckt (\hat{q})$ for any
$\covr, \hat{q} \in \covspa$, or $\Gamma_\deckt (\hat{q}) =
\Gamma_\deckt$, so that \eqref{e:compo06} becomes the usual
\zz{composition law $\Gamma_{\deckt_1 \deckt_2} = \Gamma_{\deckt_1}
\Gamma_{\deckt_2}$ and  $\Gamma$ is a unitary
representation of $\deckg$.}

\zz{The most important case of topological factors that are characters is
provided by identical particles {\it with spin}. In fact, for this case,
Assertion~\ref{a:bundle} entails the same conclusions we arrived at the end
of Section \ref{sec:covering}, even for particles with spin. To understand
how this comes about, consider the potential occurring in the Pauli
equation for $N$ identical particles with spin,
\begin{equation}\label{pauli2}
V(q) = -\mu\sum_{\canq \in q} \magnetic(\canq) \cdot
\boldsymbol{\sigma}_{\canq}
\end{equation}
on the spin bundle \eqref{spinbundle} over $\nrtre$, with
$\boldsymbol{\sigma}_{\canq}$ the vector of spin matrices acting
on the spin space of the particle at $\canq$. Clearly, the algebra
generated by $\{V(q)\}$ arising from all possible choices of the
magnetic field $\magnetic$ is $\mathrm{End}(E_q)$. Thus the only
holonomy-twisted representations that define a dynamics for all
magnetic fields are those given by a character.\footnote{In fact,
it can be shown \cite {topid2} that the only holonomy-twisted
representations for a magnetic field $\magnetic$ that is not
parallel must be a character.}}

An example of a topological factor that is not a character is
provided by the Aharonov--Casher variant \cite{AC84} of the
Aharonov--Bohm effect, according to which a neutral spin-1/2
particle that carries a magnetic moment $\mu$ acquires a
nontrivial phase while encircling a charged wire $\cylinder $.  A
way of understanding how this effect comes about is in terms of
the non-relativistic Hamiltonian $\rawH$ based on a nontrivial
connection $\nabla = \nabla_\mathrm{trivial} - \tfrac{i\mu}{\hbar}
\electric \times \boldsymbol{\sigma}$ on the vector bundle $\RRR^3
\times \CCC^2$.  Suppose the charge density $\varrho(\canq)$ is
invariant under translations in the direction $\boldsymbol{e}\in
\RRR^3$, $\boldsymbol{e}^2=1$ in which the wire is oriented. Then
the charge per unit length $\lambda$ is given by the integral
\begin{equation} \lambda = \int_D \varrho(\canq)\, dA
\end{equation} over the cross-section disk $D$ in any plane
perpendicular to $\boldsymbol{e}$. The restriction of this connection,
outside of $\cylinder $, to any plane $\Sigma$ orthogonal to the wire turns
out to be flat\footnote{The curvature is $\Omega = d_\mathrm{trivial}
\boldsymbol{\omega} + \boldsymbol{\omega} \wedge \boldsymbol{\omega} $ with
$\boldsymbol{\omega} = -i\frac{\mu}{\hbar} \electric \times
\boldsymbol{\sigma}$. The 2-form $\Omega$ is dual to the vector
$\nabla_\mathrm{trivial} \times\boldsymbol{\omega} + \boldsymbol{\omega}
\times \boldsymbol{\omega} =i\frac{\mu}{\hbar}
(\nabla\cdot\electric)\boldsymbol{\sigma} -
i\frac{\mu}{\hbar}(\boldsymbol{\sigma}\cdot\nabla)\electric - 2i
(\frac{\mu}{\hbar})^2(\boldsymbol{\sigma} \cdot \electric) \electric.$
Outside the wire, the first term vanishes and, noting that
$\electric\cdot\boldsymbol{e} =0,$ the other two terms have vanishing
component in the direction of $\boldsymbol{e}$ and thus vanish when
integrated over any region within an orthogonal plane.} so that its
restriction to the intersection $\Q$ of $\RRR^3\setminus\cylinder$ with the
orthogonal plane can be replaced, as in the Aharonov--Bohm case, by the
trivial connection if we introduce a periodicity condition on the wave
function with the topological factor
\begin{equation} \Gamma_1 = \exp \Bigl(-\frac{4\pi i\mu\lambda}{\hbar}\,
\boldsymbol{e}\cdot\boldsymbol{\sigma} \Bigr) \,.  \end{equation}
\zz{In this way} we obtain a representation $\Gamma : \fund{\Q}
\to SU(2)$ that is not given by a character.

Another example of a topological factor that is not a character
and which can be generalized to a nonabelian representation is
provided by a higher-dimensional version of the Aharonov--Bohm
effect: one may replace the \z{vector potential} in the
Aharonov--Bohm setting by a non-abelian gauge field (\`a la
Yang--Mills) \z{whose field strength (curvature) vanishes} outside
a cylinder $\cylinder$ but not inside; the value space $\sspa$
(now corresponding not to spin but to, say, quark color) has
dimension greater than one, and the difference between two wave
packets that have passed $\cylinder$ on different sides is given
in general, not by a phase, but by a unitary endomorphism $\Gamma$
of $\sspa$. In this example, involving one \x{cylinder}, the
representation $\Gamma$, though given by matrices that are not
multiples of the identity, is nonetheless abelian, since
$\fund{\Q} \cong \ZZZ$ is an abelian group.  However, when two or
more  cylinders are considered, we obtain a non-abelian
representation $\Gamma$, since when $\Q$ is $\RRR^3$ minus two
disjoint solid cylinders its fundamental group is isomorphic to
the non-abelian group $\ZZZ \ast \ZZZ$, where $\ast$ denotes the
free product of groups, generated by loops $\sigma_1$ and
$\sigma_2$ surrounding one or the other of the cylinders. One can
easily arrange that the matrices $\Gamma_{\sigma_i}$ corresponding
to loops $ \sigma_i$, $i=1,2$, fail to commute, so that $\Gamma$
is nonabelian.

Our last example involves a holonomy-twisted representation
$\Gamma$ that is not a representation in the ordinary sense.
Consider $N$ fermions, each as in the previous examples, moving in
$M=\RRR^3\setminus \cup_i\cylinder_i$, where $\cylinder_i$ are one
or more disjoint solid cylinders. More generally, consider $N$
fermions, each having 3-dimensional configuration space $M$ and
value space $W$ (which may incorporate spin or ``color'' or both).
Then the configuration space $\Q$ for the $N$ fermions is the set
${}^N\! M$ of all $N$-element subsets of $M$, \x{with universal
covering space $\covspa=\widehat {{}^N\! M} =
{\widehat{M}}^N\setminus \Delta$ with $\Delta$} the extended
diagonal, the set of points in ${\widehat{M}}^N$ whose projection
to $M^N$ lies in its coincidence set. Every diffeomorphism
$\sigma\in Cov(\widehat{{}^N\! M}, {}^N\! M)$ can be expressed as
a product
\begin{equation}\label{prod}
\sigma=p\tilde\sigma
\end{equation}
where $p \in S_N$ and $\tilde\sigma =
(\sigma^{(1)},\dots,\sigma^{(N)})\in Cov(\widehat{M},M)^N $ and these act on
$\covq=(\hat\canq_1,\dots, \hat\canq_N)$ $\in\widehat{M}^N$ as follows:
\begin{equation}\label{tildesigmaq}
\tilde\sigma\covq=(\sigma^{(1)}\hat\canq_1,\dots, \sigma^{(N)}\hat\canq_N)
\end{equation}
and
\begin{equation}\label{pq}
p\covq=(\hat\canq_{p^{-1}(1)},\dots, \hat\canq_{p^{-1}(N)}).
\end{equation}
Thus
\begin{equation}\label{sigmaq}
\sigma\covq=(\sigma^{(p^{-1}(1))}\hat\canq_{p^{-1}(1)},\dots, \sigma^{(p^{-1}(N))}\hat\canq_{p^{-1}(N)}).
\end{equation}
Moreover, the representation (\ref{prod}) of $\sigma$ is unique. Thus,
since
\begin{equation}\label{sdp}
\sigma_1\sigma_2=p_1\tilde\sigma_1p_2\tilde\sigma_2=(p_1p_2)(p_2^{-1}\tilde\sigma_1p_2\tilde\sigma_2)
\end{equation}
with
$p_2^{-1}\tilde\sigma_1p_2=(\sigma_1^{(p_2(1))},\dots,\sigma_1^{(p_2(N))})\in Cov(\widehat{M},M)^N$,
we find that
$Cov(\widehat{{}^N\! M}, {}^N\! M)$ is a semidirect product of $S_N$ and
$Cov(\widehat{M},M)^N$, with product
given by
\begin{equation}\label{sprod}
\sigma_1\sigma_2=(p_1,\tilde\sigma_1)(p_2,\tilde\sigma_2)=(p_1p_2,p_2^{-1}\tilde\sigma_1 p_2\tilde\sigma_2).
\end{equation}

Wave functions for the $N$ fermions are sections of the lift $\widehat E$ to
$\covspa$ of the bundle $E$ over $\Q$ with fiber
\begin{equation}\label{nw}
E_q=\bigotimes_{\canq \in q} W
\end{equation}
and  (nontrivial) connection inherited from the trivial connection
on $M\times W$. If the dynamics for $N=1$ involves wave functions on
$\widehat{M}$ obeying (\ref {e:percon06}) with topological factor
$\Gamma_\deckt(\hat{\boldsymbol{q}})=\Gamma_\deckt$ given by a
unitary \x{representation of $\fund{M}$ (i.e., independent of
$\hat{\boldsymbol{q}}$), then} the $N$ fermion wave \x{function
obeys} (\ref{e:percon06}) with topological factor

\begin{equation}\label{biggamma}
\Gamma_{\sigma}(\covq)=\mathrm{sgn}(p)\bigotimes_{\canq \in
\pi(\covq)}\Gamma_{\sigma^{(i_{\covq}(\canq))}}
\equiv \mathrm{sgn}(p)\Gamma_{\tilde\sigma}(\covq)
\end{equation}
where for  $\covq=(\hat\canq_1,\dots,\hat\canq_N), \
\pi(\covq)=\{\pi_M(\hat\canq_1),\dots,\pi_M(\hat\canq_N)\}$ and
$i_{\covq}(\pi_M(\hat\canq_j))=j$. Since

\begin{equation}\label{prod2}
\Gamma_{\tilde\sigma_1 \tilde\sigma_2}(\covq) = \Gamma_{\tilde\sigma_1}
(\covq) \, \Gamma_{\tilde\sigma_2} (\covq)
\end{equation}
we find, using (\ref{sprod}) and  (\ref{prod2}), that

\begin{subequations}\label{htr}
\begin{align}
\Gamma_{\sigma_1 \sigma_2}(\covq)&=
\mathrm{sgn}(p_1p_2)\Gamma_{p_2^{-1}\tilde\sigma_1
p_2\tilde\sigma_2}(\covq)\\ &=\mathrm{sgn}(p_1)\Gamma_{p_2^{-1}\tilde\sigma_1
p_2}(\covq)\mathrm{sgn}(p_2)\Gamma_{\tilde\sigma_2}
(\covq)\\ &=P_2\Gamma_{\sigma_1}(\covq)P_2^{-1}\Gamma_{\sigma_2}
(\covq),
\end{align}
\end{subequations}
which agrees with (\ref{twistedrep}) since the holonomy on the bundle $E$
is given by permutations $P$ acting on the tensor product (\ref{nw}).

\section{Conclusions}
\label{sec:conclusions}

We have \zz{investigated} the possible quantum theories on a topologically
nontrivial configuration space $\Q$ from the point of view of Bohmian
mechanics, which is fundamentally concerned with the motion of matter
in physical space, represented by the evolution of a point in
configuration space.

Our goal was \zz{to find} all Bohmian dynamics in \gencon, where the wave
functions may be sections of a \herm\ vector bundle $E$. What ``all''
Bohmian dynamics means is not obvious; we have followed one approach to
what it can mean; other approaches will be described in \zz{future}
works. The present approach uses \z{ wave functions $\psi$} that are
defined on the universal covering space $\covspa$ of \gencon\ and satisfy a
periodicity condition ensuring that the Bohmian velocity vector field on
$\covspa$ defined in terms of $\psi$ can be projected to \gencon. We have
arrived in this way at a natural class of Bohmian dynamics beyond the
immediate Bohmian dynamics.  Such a dynamics is defined by a potential and
some information encoded in ``topological factors,'' which form either a
character (one-dimensional unitary representation) of the fundamental group
of \z{the} configuration space, $\fund{\Q}$, or a more general
algebraic-geometrical object, a \zz{holonomy-twisted representation} $\Gamma$.  Only
those dynamics associated with characters are compatible with \emph{every}
potential, as one would desire for \z{what} could be \z{considered} a
version of quantum mechanics in $\Q$.  We have thus arrived at the known
fact that for every character of $\fund{\Q}$ there is a version of quantum
mechanics in $\Q$. A consequence, which will be discussed in detail in a
sister paper \cite{topid2}, is the symmetrization postulate for identical
particles. These different quantum theories emerge naturally when one
contemplates the possibilities for defining a Bohmian dynamics in $\Q$.

\section*{Acknowledgments}

We thank Kai-Uwe Bux (Cornell University), Frank Loose
(Eberhard-Karls-Universit\"at T\"ubingen, Germany) and Penny Smith
(Lehigh University) for helpful discussions.

R.T.\ gratefully acknowledges support by the German National Science
Foundation (DFG) through its Priority Program ``Interacting Stochastic
Systems of High Complexity'', by INFN, and by the European Commission
through its 6th Framework Programme ``Structuring the European
Research Area'' and the contract Nr. RITA-CT-2004-505493 for the
provision of Transnational Access implemented as Specific Support
Action.  The work of S.~Goldstein was supported in part by NSF Grant
DMS-0504504. N.Z.\ gratefully acknowledges support by INFN.

%Finally,
We appreciate the hospitality that some of us have enjoyed, on more
than one occasion, at the Mathematisches Institut of
Ludwig-Maximilians-Universit\"at M\"unchen (Germany), the
Dipartimento di Fisica of Universit\`a di Genova (Italy), the
Institut des Hautes \'Etudes Scientifiques in Bures-sur-Yvette
(France), and the Mathematics Department of Rutgers University
(USA).

Finally we would like to thank an anonymous referee for helpful
criticisms on an earlier version of this article.

\end{document}